\documentclass[12pt]{spieman} 
\usepackage{amsmath,amsfonts,amssymb}
\usepackage{graphicx}
\usepackage{setspace}
\usepackage{tocloft}
\usepackage{lineno}

\title{Construction and Testing of a Common Mode Choke for Cryogenic Detector Pre-Amplifiers}

\author[a]{Mathias Richerzhagen}
\author[a]{Joshua Hopgood}
\affil[a]{European Southern Observatory (ESO), Karl-Schwarzschild-Str. 2, 85748 Garching bei München, Germany}

\cftpagenumbersoff{figure}
\cftpagenumbersoff{table} 
\begin{document} 
\maketitle

\begin{abstract}
Common-mode choke inductors are useful tools for resolving grounding issues in large detector systems. Using inductive components on cryogenic pre-amplifier boards has so far been prevented by the poor low-temperature performance of common ferrite materials such as NiZn and MnZn. Recently developed nanocrystalline and amorphous ferrite materials promise improved performance up to the point where using magnetics at liquid mitrogen temperatures becomes feasible. This research applies the work of Yin et al.\cite{9595353} on characterizing ferrite materials by constructing and testing a common mode choke inductor for use on detector pre-amplifiers for the ELT first generation instruments.
\end{abstract}


\begin{spacing}{2} 

\section{Introduction}
\label{sect:intro} 
Common mode chokes are a well-understood instrument for improving immunity to electromagnetic disturbances as described among many in Ref.~[\citenum{9259}]. As passive components consisting of two windings on a common ferromagnetic core, chokes provide high impedance for common mode signals while having significantly lower impedance to differential signals. These properties make them efficient at preventing the coupling of common mode disturbance signals into electronic circuits through power or data lines, thus reducing the occurrence of a class of artifacts summarized as ``Grounding Issues'' during integration and testing. In cryogenic detector systems, the use of common mode chokes on power and signal inputs may be especially advantageous since the detector and associated pre-amplifier board are typically located far away from other control electronics (i.e. the detector controller). For the ELT first generation instruments METIS \cite{10.1117/12.2056468}, HARMONI \cite{10.1117/12.2562144} and MICADO\cite{10.1117/12.2233047} the warm control electronics cabinet and cold pre-amplifier are separated by cables of up to 5m length. Ideally, differential power and signal inputs of a cryogenic pre-amplifier would be common mode filtered at the pre-amplifier. According to Ref.~[\citenum{8854889}], the use of common magnetic core materials is not feasible in cryogenic systems due to the significant loss in magnetic permeability of the core at liquid nitrogen temperatures. Recently-developed nanocrystalline and amorphous core materials have been shown to perform much better at low temperatures as evaluated in Ref.~[\citenum{9595353}]. This research aims at applying the results of Ref.~[\citenum{9595353}] to construct a common mode choke inductor suitable for cryogenic detector systems and comparing its performance to classic NiZn or MnZn based cores. The inductor is characterized in terms of its impedance over frequency curve and not its magnetic properties for easier use in electronics engineering design documentation.

\section{Method}
\subsection{Construction}
Four common mode choke samples are constructed on commercially available toroidal cores.
\begin{figure}[ht]
\begin{center}
\begin{tabular}{c}
\includegraphics[height=4cm]{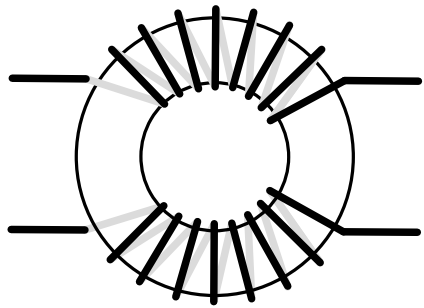}
\end{tabular}
\end{center}
\caption 
{ \label{fig:construction}
Segmented Winding Common Mode Choke } 
\end{figure} 

\begin{figure}[ht]
\begin{center}
\begin{tabular}{c}
\includegraphics[height=5.5cm]{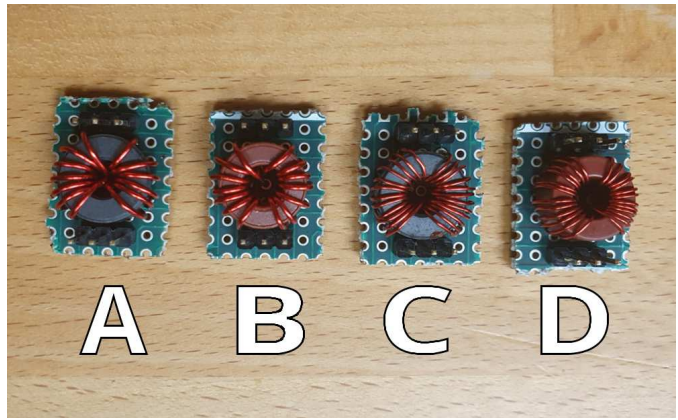}
\end{tabular}
\end{center}
\caption 
{ \label{fig:samples}
Common Mode Choke Samples } 
\end{figure} 

\begin{table}[ht]
\caption{Common Mode Choke Sample Parameters} 
\label{tab:samples}
\begin{center}
\begin{tabular}{|l|l|l|} 
\hline
\rule[-1ex]{0pt}{3.5ex} Sample & Core & Windings \\
\hline\hline
\rule[-1ex]{0pt}{3.5ex} A & ½ Würth 74270113 NiZn & 2x 6 Windings AWG22 Magnet Wire \\
\hline
\rule[-1ex]{0pt}{3.5ex} B & VAC L2009-W914 Nanocrystalline & 2x 6 Windings AWG22 Magnet Wire \\
\hline
\rule[-1ex]{0pt}{3.5ex} C & ½ Würth 74270113 NiZn & 2x 12 Windings AWG26 Magnet Wire \\
\hline
\rule[-1ex]{0pt}{3.5ex} D & VAC L2009-W914 Nanocrystalline & 2x 12 Windings AWG26 Magnet Wire \\
\hline
\end{tabular}
\end{center}
\end{table} 

 Samples are constructed using nanocrystalline and traditional NiZn based core materials and placed on a test coupon made from perforated circuit board. For the NiZn core, a larger core is split in half to achieve a comparable magnetic cross section to the nanocrystalline core. Windings are added manually using a segmented winding pattern as shown in Figure~\ref{fig:construction}. A bifilar winding pattern could be chosen as an alternative, but the segmented technique allows for some spatial separation between the two windings making the effect of an insulation failure less severe since a winding-to-winding short circuit is less likely. The resulting inductor samples are shown in Figure~\ref{fig:samples} with construction parameters listed in Table~\ref{tab:samples}.

\subsection{Testing}

\begin{figure}[h]
\begin{center}
\begin{tabular}{c}
\includegraphics[width=\textwidth]{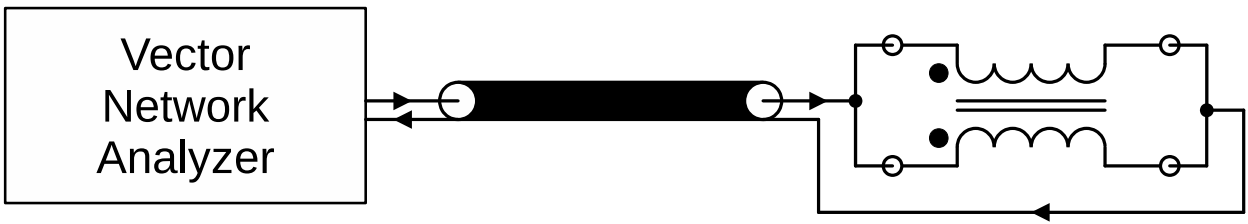}
\end{tabular}
\end{center}
\caption 
{ \label{fig:setup_comm}
Common Mode Test Setup } 
\end{figure} 
\begin{figure}[h]
\begin{center}
\begin{tabular}{c}
\includegraphics[width=\textwidth]{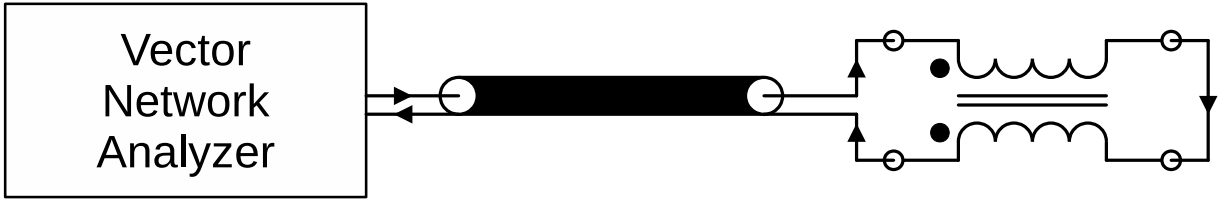}
\end{tabular}
\end{center}
\caption 
{ \label{fig:setup_diff}
Differential Mode Test Setup } 
\end{figure} 

\begin{figure}[h]
\begin{center}
\begin{tabular}{c}
\includegraphics[height=5.5cm]{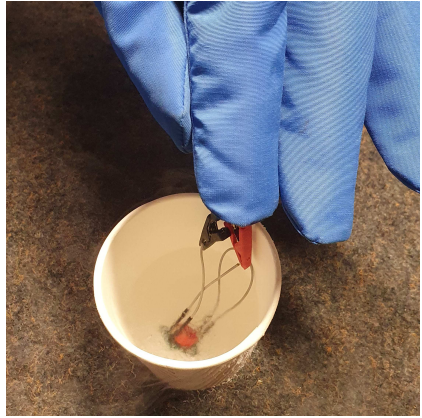}
\end{tabular}
\end{center}
\caption 
{ \label{fig:setup}
Liquid Nitrogen Immersion Setup } 
\end{figure}

The test samples are connected to a vector network analyser (OMICRON Lab - Bode 100) configured in one-port impedance measurement mode as shown in Figure~\ref{fig:setup_comm} and Figure~\ref{fig:setup_diff} to measure common mode and differential impedance over frequency both at room temperature (approximately 293K) and liquid nitrogen temperature (77K at standard pressure). The chosen vector network analyser is capable of directly plotting the impedance magnitude trace over frequency. The supported frequency range is up to 40MHz. For the cold temperature test, the entire sample is manually submersed in liquid nitrogen as shown in Figure~\ref{fig:setup} until steady state is reached.

\section{Results}
\begin{figure}[h]
\begin{center}
\begin{tabular}{c}
\includegraphics[width=\textwidth]{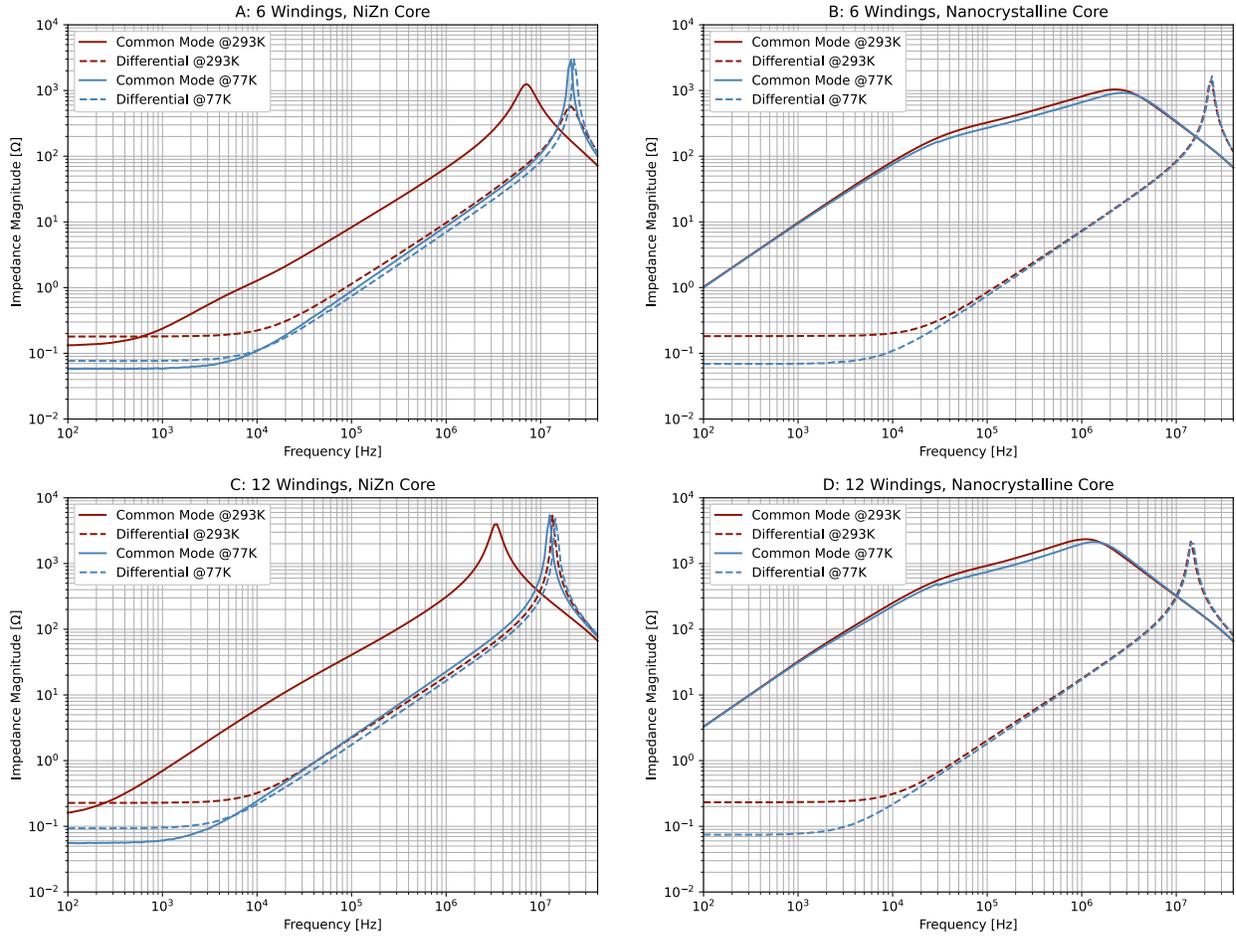}
\end{tabular}
\end{center}
\caption 
{ \label{fig:results}
Measurement Results } 
\end{figure}

Test results are presented as Bode plots of impedance magnitude over frequency in Figure~\ref{fig:results} since these plots are often found in the data sheets of commercial common mode chokes which allows comparison with other parts.

It is observed that at room temperature, all samples show the typical impedance curve of common mode choke inductors. In differential mode the impedance is clearly lower than in common mode since magnetic flux induced by both windings cancels out. A clear resonance peak caused by the parasitic winding capacitance is present. In common mode operation, the nanocrystalline chokes exhibit higher impedance than the traditional core materials. This is in line with expectations compared to commercially available nanocrystalline inductors. The performance advantage is offset by higher cost. At cryogenic temperature it becomes apparent that the common mode impedance of the traditional NiZn cores degrades until the point where it is almost indistinguishable from the differential mode impedance. This indicates that the core material is no longer effective, confirming the result stated in Ref.~[\citenum{9595353}]. The nanocrystalline cores maintain performance except for an insignificant decrease in impedance. It is also observed that the DC resistance visible in the low frequency range of the differential impedance curves decreases with temperature, as expected. 

\section{Discussion}
The experiments show that common mode choke inductors based on commercially-available nanocrystalline core materials are suitable for operation at liquid nitrogen temperatures since the core material maintains its magnetic properties at cold temperature. Commercial inductors based on similar core materials may be considered for application in detector control systems operating at cryogenic temperatures.

The tested common mode chokes are most suitable for decoupling the preamp power supply lines due to the chosen core size and winding wire gauge. It may be advantageous to also common-mode filter signal inputs in the future. At the time of writing, no small-size nanocrystalline cores were available for purchase that allow manual construction of sufficiently small data line common mode chokes. With the cooperation of a commercial magnetics manufacturer, it may be possible to adapt smaller signal line common mode chokes, possibly in surface mount technology form factors, to nanocrystalline core materials. Due to the general performance advantage of nanocrystalline cores even at room temperature it is expected that magnetic components using this technology will soon become more widely available.

Long-term reliability remains to be demonstrated, but it is promising that none of the cores or windings were damaged by repeated immersion in liquid nitrogen during this work. In general, it is advisable to cool down ferrite materials slowly due to the risk of cracking. To prevent contamination of the cryostat in case a core does fracture it is also advisable to use epoxy coated cores to keep fragments from separating.

Before inclusion in a scientific detector system, other parameters of the chokes such as winding resistance, DC bias current, thermal effects, and self-resonance need to be characterised. Consideration of these parameters should be part of the standard design procedure for a common mode choke and are expected to be unlikely to change at cold temperatures based on the test results presented here.


\bibliography{report} 
\bibliographystyle{spiejour} 


\vspace{2ex}\noindent\textbf{Mathias Richerzhagen} is a detector electronics engineer at the European Southern Observatory. He received his Engineering Diploma from RWTH Aachen University in 2012. His current research includes development of the detector controller for the ELT as well as some work in cryo-electronics.

\vspace{1ex}
\noindent Biographies and photographs of the other authors are not available.

\listoffigures
\listoftables

\end{spacing}
\end{document}